\documentclass[twocolumn]{aastex62}

\usepackage{graphicx}	% Including figure files
\usepackage{amsmath}	% Advanced maths commands
\usepackage{amssymb}	% Extra maths symbols
\usepackage{xcolor}	    % Color

\shorttitle{Elemental Abundances of LIRG Arp\,299}
\shortauthors{Mao et al.}
\begin{document}

\title{Elemental Abundances of the Hot Atmosphere of Luminous Infrared Galaxy Arp\,299}

\correspondingauthor{Junjie Mao}
\email{jmao2018@hiroshima-u.ac.jp}

\author[0000-0001-7557-9713]{Junjie Mao}
\affiliation{Department of Physics, Hiroshima University, 1-3-1 Kagamiyama, HigashiHiroshima, Hiroshima 739-8526, Japan}
\affiliation{Department of Physics, University of Strathclyde, Glasgow G4 0NG, UK}
\affiliation{SRON Netherlands Institute for Space Research, Sorbonnelaan 2, 3584 CA Utrecht, the Netherlands}

\author[0000-0002-5683-822X]{Ping Zhou}
\affiliation{School of Astronomy and Space Science, Nanjing University, Nanjing 210023, PR China}

\author[0000-0002-9714-3862]{Aurora Simionescu} 
\affiliation{SRON Netherlands Institute for Space Research, Sorbonnelaan 2, 3584 CA Utrecht, the Netherlands}
\affiliation{Leiden Observatory, Leiden University, PO Box 9513, 2300 RA Leiden, The Netherlands}
\affiliation{Kavli Institute for the Physics and Mathematics of the Universe (WPI), The University of Tokyo, Kashiwa, Chiba 277-8583, Japan}

\author[0000-0002-3886-1258]{Yuanyuan Su} 
\affiliation{Department of Physics and Astronomy, University of Kentucky, 505 Rose Street, Lexington, KY, 40506, USA}

\author[0000-0002-0921-8837]{Yasushi Fukazawa}
\affiliation{Department of Physical, Hiroshima University, 1-3-1 Kagamiyama, Higashi-Hiroshima, Hiroshima 739-8526, Japan}
\affiliation{Hiroshima Astrophysical Science Center, Hiroshima University, 1-3-1 Kagamiyama, Higashi-Hiroshima, Hiroshima 739–8526, Japan}
\affiliation{Core Research for Energetic Universe, Hiroshima University, 1-3-1 Kagamiyama, Higashi-Hiroshima, Hiroshima 739–8526, Japan}

\author[0000-0001-9911-7038]{Liyi Gu} 
\affiliation{SRON Netherlands Institute for Space Research, Sorbonnelaan 2, 3584 CA Utrecht, the Netherlands}
\affiliation{RIKEN High Energy Astrophysics Laboratory, 2-1 Hirosawa, Wako, Saitama 351-0198, Japan}

\author[0000-0003-1949-7005]{Hiroki Akamatsu}
\affiliation{SRON Netherlands Institute for Space Research, Sorbonnelaan 2, 3584 CA Utrecht, the Netherlands}

\author[0000-0001-8812-8284]{Zhenlin Zhu} 
\affiliation{SRON Netherlands Institute for Space Research,  Sorbonnelaan 2, 3584 CA Utrecht, the Netherlands}
\affiliation{Leiden Observatory, Leiden University, PO Box 9513, 2300 RA Leiden, The Netherlands}

\author[0000-0002-2697-7106]{Jelle de Plaa} 
\affiliation{SRON Netherlands Institute for Space Research, Sorbonnelaan 2, 3584 CA Utrecht, the Netherlands}

\author[0000-0002-7031-4772]{Fran\c{c}ois Mernier} 
\affiliation{European Space Agency (ESA), European Space Research and Technology Centre (ESTEC) \\ Keplerlaan 1, 2201 AZ Noordwijk, The Netherlands}
\affiliation{SRON Netherlands Institute for Space Research,  Sorbonnelaan 2, 3584 CA Utrecht, the Netherlands}

\author[0000-0001-5540-2822]{Jelle S. Kaastra} 
\affiliation{SRON Netherlands Institute for Space Research, Sorbonnelaan 2, 3584 CA Utrecht, the Netherlands}
\affiliation{Leiden Observatory, Leiden University, PO Box 9513, 2300 RA Leiden, The Netherlands}

%\author{A. N. Other}



\begin{abstract}
Hot atmospheres of massive galaxies are enriched with metals. Elemental abundances measured in the X-ray band have been used to study the chemical enrichment of supernova remnants, elliptical galaxies, groups and clusters of galaxies. Here we measure the elemental abundances of the hot atmosphere of luminous infrared galaxy Arp\,299 observed with \textit{XMM-Newton}. To measure the abundances in the hot atmosphere, we use a multi-temperature thermal plasma model, which provides a better fit to the Reflection Grating Spectrometer data. The observed Fe/O abundance ratio is subsolar, while those of Ne/O and Mg/O are slightly above solar. Core-collapse supernovae (SNcc) are the dominant metal factory of elements like O, Ne, and Mg. We find some deviations between the observed abundance patterns and theoretical ones from a simple chemical enrichment model. One possible explanation is that massive stars with $M_{\star}\gtrsim23-27~M_{\odot}$ might not explode as SNcc and enrich the hot atmosphere. This is in accordance with the missing massive SNcc progenitors problem, where very massive progenitors $M_{\star}\gtrsim18~M_{\odot}$ of SNcc have not been clearly detected. It is also possible that theoretical SNcc nucleosynthesis yields of Mg/O yields are underestimated. 
\end{abstract}
% words count: 190



\section{Introduction} 
\label{sct:intro}
Hot atmospheres of massive galaxies are enriched with metals \citep{wer19}. Metals are mainly produced by stars before and during their splendid death as supernovae (SN). Generally speaking, core-collapse supernovae (SNcc) of massive $(M_*\gtrsim8~M_{\odot})$ stars are the main metal factory of elements like O, Ne, and Mg \citep{nom13}. Type-Ia supernovae (SNIa) dominate the enrichment of the Fe-peak elements like Fe and Ni \citep{hit17}. Low- and intermediate-massive stars $(M_*\lesssim7~M_{\odot})$ in the asymptotic giant branch (AGB) phase contribute most to light elements like N \citep{kob06,wer06,mao19}. 

In chemical enrichment models, it is common to assume that massive stars from $\sim10~M_{\odot}$ to $\gtrsim40-50~M_{\odot}$ can explode as SNcc and enrich the surrounding environment \citep[e.g.,][]{kob06,san06,dpl07,mer16,mao19}. If the SNcc progenitor does not contain metals in its atmosphere (i.e. $Z_{\rm init}=0$), very massive stars up to $M_*=140~M_{\odot}$ can also contribute to the chemical enrichment via SNcc \citep{nom13}. It is still debated whether very massive stars with higher initial metalicity could make successful SN explosions \citep{heg03}.
% Fig. 1 of kob06 

From the observational perspective, there is a missing massive SNcc progenitor problem as reviewed by \citet{sma15}. Massive stars, especially those with $M_{\star}\gtrsim18~M_{\odot}$ might explode as a hypernova (HN), explode as a faint supernova ejecting a small amount of heavy elements, or directly form a black hole without a visible supernova \citep{nom13}. If the bulk of massive stars above a certain mass limit do not explode as SNcc or HN, we expect to see mismatches between the observed and theoretical abundance ratios.

Elemental abundances measured in the X-ray band have been used to decode the chemical enrichment of supernova remnants \citep[e.g.,][]{zho18}, elliptical galaxies, groups and clusters of galaxies \citep[e.g.,][]{hit17,mer18,sim19,mao19}. Here, we present the elemental abundances of the hot atmosphere of a starburst galaxy, which form stars intensively. 
% see also zho19,zho21

\object{Arp\,299}, also known as \object{NGC 3690}\footnote{IC 694 is a small elliptical or lenticular galaxy about an arcmin northwest, which is not part of Arp\,299.}, is a pair of interacting galaxies with a star-formation rate of $119~M_{\odot}~{\rm yr^{-1}}$ \citep{smi18} at the distance of $48.5\pm3.4$~Mpc \citep{jam14}. It is one of the most powerful nearby starburst galaxies \citep{alo09}. With $L_{\rm IR}=6.3\times10^{14}~L_{\odot}$ \citep{per11}, it is qualified as a luminous infrared galaxy \citep{san96}. From 1992 to 2010, seven SNe have been recorded in Arp\,299 \citep{and11}. Six of them are identified as SNcc and one is unclassified \citep{and11}. According to the Open Supernova Catalog \citep{gui17}\footnote{https://sne.space/}, three more SNe have been found since 2010: SN2018lrd, SN2019lqo, and SN2020fkb. All of these three are SNcc. Nevertheless, due to the obscuration by large amounts of dust in Arp\,299, \citet{mat12} argue that $\sim$80\% of the SNe are estimated to be missed by observations in the optical band. 

Arp\,299 hosts a hot atmosphere that dominates the soft X-ray spectrum below 2~keV \citep[e.g.,][]{ana16}. About 20 ultraluminous X-ray sources \citep{zez03,ana16} and at least one Compton-thick nucleus \citep{pta15} dominate the hard X-ray spectrum at $2-10$ keV and $>10$ keV, respectively. Previous X-ray studies do not focus on accurate abundance measurement of O, Ne, and Mg, which are mainly enriched by SNcc. Here, we measure the elemental abundances of the hot atmosphere of Arp\,299 with the high-quality X-ray spectra obtained with \textit{XMM-Newton}. We compare the observed abundance ratios with various chemical enrichment models. These models include different SN yields, the initial mass functions (IMFs), the initial metallicities and the mass ranges of the SNcc progenitors, and mixing fractions of HN or SNIa. 

A supplementary package is available at Zenodo\footnote{https://doi.org/10.5281/zenodo.5148020}. This package includes data and scripts used to create the figures presented in this paper.

% word count: 594

\section{\textit{XMM-Newton} observations}
\label{sct:obs_dr}
We use all the X-ray spectra observed with \textit{XMM-Newton} (Table~\ref{tbl:obs_log}), including three new observations obtained in Cycle 19 (PI: J. Mao). The data are reduced following the standard procedure using \textit{XMM-Newton} Science Analysis Software (SAS) v19.0. We construct a time-averaged X-ray spectrum with the high-resolution soft X-ray spectra ($7.5-27$~\AA) obtained with Reflection Grating Spectrometer \citep[RGS,][]{dhe01} and the relatively low-resolution hard X-ray spectra ($1.6-10.5$~\AA) obtained with the positive-negative (pn) junction CCD (charge-coupled device) camera of the European Photon Imaging Camera (EPIC). 

\begin{deluxetable}{c|cc}
\tablecaption{\textit{XMM-Newton} observation log. 
\label{tbl:obs_log}}
\tablehead{
\colhead{Date} & \colhead{ObsID} & \colhead{Effective exposures}
}
%\colnumbers
\startdata
2001-05-06 & 0112810101 & 9.6~ks (RGS), 13.0 ks (pn) \\
2011-12-15 & 0679381101 & 9.8~ks (RGS), 6.4 ks (pn) \\
2020-05-08 & 0861250101 & 42.9~ks (RGS), 39.3 ks (pn) \\
2020-05-22 & 0861250201 & 34.9~ks (RGS), 26.3 ks (pn) \\
2020-11-22 & 0861250301 & 35.7~ks (RGS), 11.6 ks (pn) \\
\noalign{\smallskip} 
\enddata
%\tablenotetext{a}{TBA}
%\tablecomments{TBA}
\end{deluxetable}

\begin{figure*}[!t]
\centering
\includegraphics[width=.9\hsize, trim={0.5cm 0.cm 0.5cm .cm}, clip]{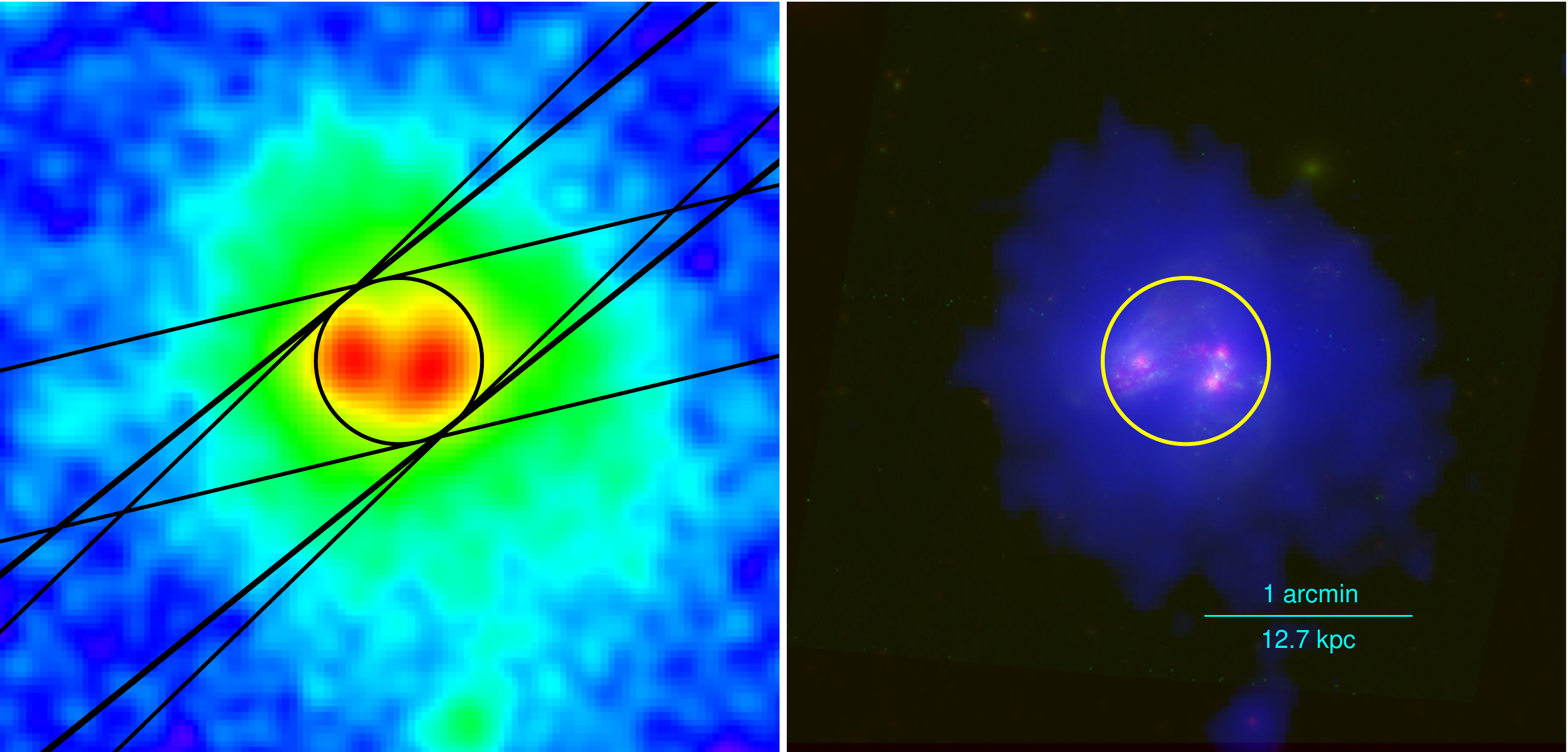}
\caption{X-ray to IR image of Arp\,299. \textit{Left}: \textit{XMM-Newton} soft X-ray ($0.3-2$~keV) image with the RGS (rectangles) and EPIC/pn (circle) source regions. The three new observations in 2020 (Table~\ref{tbl:obs_log}) have similar RGS roll angles so that one rectangular box appears thick. \textit{Right}: The true color image of Arp\,299 with soft X-ray in purple, optical (\textit{Hubble}, $814$~nm) in green, and infrared (\textit{Spitzer}, 3.6~$\mu$m) in red. }
\label{fig:xray2IR_zoom} 
\end{figure*}
% 2001-05-06: PA = 314
% 2011-12-15: PA = 103
% 2020-05-08: PA = 309
% 2020-05-22: PA = 309
% 2020-11-22: PA = 129 

Lightcurves from the CCD9 of RGS and the $10-12$~keV energy band of EPIC/pn are created with a time bin of 100~s. Time intervals contaminated by background soft-proton flares are identified as those above the $3\sigma$ level and excluded from the following data reduction. As shown in the left panel of  Figure~\ref{fig:xray2IR_zoom}, we extract RGS spectra in rectangular regions centered on the source with the width along the cross-dispersion direction equals $\sim48$~arcsec (i.e., 90~\% of the point spread function). We extract EPIC/pn spectra in a circular region with a radius of 24~arcsec. The RGS modeled background and pn local background spectra are subtracted. All the RGS spectra are combined, as well as the EPIC/pn spectra. By matching the flux in the common $7.5-10.5$~\AA\ wavelength range, the EPIC/pn spectra are scaled by 0.70 with respect to RGS to account for the different instrument normalization and the different aperture (the left panel of Figure~\ref{fig:xray2IR_zoom}). The right panel of Figure~\ref{fig:xray2IR_zoom} shows a true color image of Arp\,299. The soft X-ray ($0.3-2$~keV) image (in purple) is created with the \textit{XMM-Newton} \textit{image}\footnote{https://www.cosmos.esa.int/web/xmm-newton/images} script. The optical image (in green) is taken with the Advanced Camera for Surveys Wide Field Channel (F814W filter, 814~nm) aboard \textit{Hubble Space Telescope}. The infrared image (in red) is taken with the Infrared Array Camera (Channel 1, 3.6~$\mu$m) aboard \textit{Spitzer} Space Telescope. 

% word count: 313

\section{Spectral analysis}
\label{sct:mo}
For the spectral analysis, we use the SPEX code \citep[][v3.06.00]{kaa20}, which includes the most recent atomic data for the Fe-L complex \citep{gu20}. We use $C$-statistics \citep{kaa17} and statistical uncertainties are quoted at the 68\% confidence level. With $H_0=70~{\rm km~s^{-1}~Mpc^{-1}}$, $\Omega_M=0.3$, and $\Omega_{\Lambda}=0.7$, the redshift of the target is 0.0112 and a distance scale of 12.7 kpc/arcmin. 

The hot atmosphere is modeled with multi-temperature thermal plasma models. A power-law component (denoted as PL) is included to account for the point sources. Three Fe K lines with rest-frame energies of 6.4~keV, 6.7~keV, and 6.97~keV are also included. These Fe K lines are associated with the nucleus of NGC 3690, X-ray binaries and supernova remnants \citep{pta15,ana16}. Since the power-law component dominates the flux above $\gtrsim2$~keV, the absolute abundance of the thermal plasma with respect to hydrogen cannot be well determined. We set the reference element to oxygen. The proto-solar abundance table of \citet{lod09} is used\footnote{The abundances denoted as $X$ and $X/O$ designate $(X/{\rm H})_{\rm gas}/(X/{\rm H})_{\odot}$ and $(X/{\rm O})_{\rm gas}/(X/{\rm O})_{\odot}$, respectively. They correspond to the ratios of $X/{\rm H}$ and $X/{\rm O}$ in the hot atmosphere to those in the proto-solar abundance table of \citet{lod09}.}. Both the thermal and power-law components are corrected for absorptions of both the host galaxy and Milky Way. For the Galactic absorption, we use $N_{\rm H}^{\rm MW}=9.35\times10^{19}~{\rm cm^{-2}}$ \citep{wil13}. Similar to \citet{smi18}, we use different hydrogen column densities for the thermal and power-law components for the host galaxy absorption. 

\begin{figure*}[!t]
\centering
\includegraphics[width=\hsize, trim={0.5cm 0.cm 1.cm .cm}, clip]{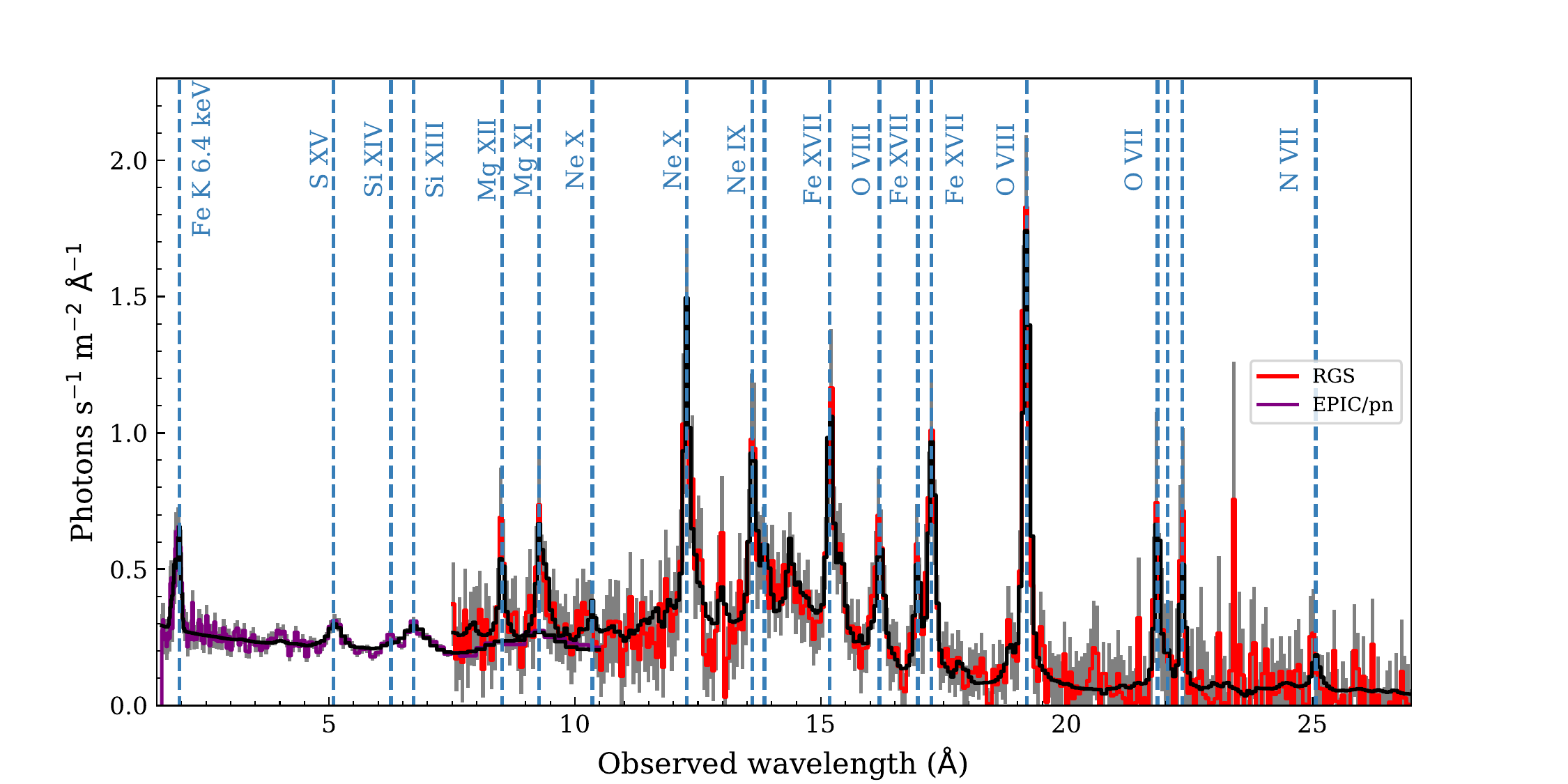}
\caption{The best-fit model (Model M4 in Table~\ref{tbl:fit_pars}) to the X-ray spectrum of Arp\,299 observed with \textit{XMM-Newton}. The high-resolution RGS spectrum is shown in red, while the EPIC/pn spectrum is shown in purple. The $1\sigma$ uncertainties are shown in grey. The RGS spectrum is rebinned here for clarity. Vertical dashed lines in blue mark key diagnostics emission lines in the spectrum. }
\label{fig:plot_dnma_01to20}
\end{figure*}

We ran a total of eight spectral fittings considering (1) various differential emission measure distributions, (2) mimicking the non-equilibrium scenario, (3) varying power-law photon indices, and (4) coupling the Fe and Ni abundances. Detailed results are presented in Appendix~\ref{sct:spec_mo}. In Figure~\ref{fig:plot_dnma_01to20}, we show the best-fit model to the observed X-ray spectrum of Arp\,299. Most of the previous studies \citep{huo04,ana16,smi18} described the hot atmosphere as a single-temperature thermal (i.e., isothermal) component. Our best-fit results favor a multi-temperature description of the hot atmosphere. This should better reflect the multi-temperature nature of the hot atmospheres of starburst galaxies \citep{str00}. 
% word count: 304

\section{Metal abundances}
Accurate abundance measurement of individual elements is essential to understand the chemical enrichment process of the hot atmosphere of Arp\,299. In the present work, we measure the abundance ratios of N/O$=0.70_{-0.29}^{+0.38}$, Ne/O$=1.16\pm0.21$, Mg/O$=1.27\pm0.26$, Si/O$=0.97\pm0.23$, S/O$=1.78\pm0.57$, Fe/O$=0.40_{-0.05}^{+0.06}$, and Ni/O$=0.66_{-0.27}^{+0.31}$. Our measurement takes advantage of the resolving power of the RGS spectra for emission lines of N, O, and Fe-L. Previous studies either do not measure individual elements separately or only cover a few elements. \citet{huo04} measured merely the Fe abundance ($0.12_{-0.05}^{+0.21}$) of the thermal plasma model MEKAL \citep{mew95}, which does not contain the state-of-the-art atomic data. \citet{ana16} used VAPEC and measure the abundances of Ne ($1.24_{-0.29}^{+0.34}$), Mg ($1.14_{-0.20}^{+0.25}$), and Fe ($0.26_{-0.03}^{+0.04}$). \citet{smi18} used VMEKAL and measured merely the $\alpha/{\rm Fe}(=3.25\pm2.48)$ abundance ratio. 

We also note that the abundance ratios measured in the X-ray band are more robust than the absolute abundance with respect to hydrogen, especially for relatively cool thermal plasmas with its weak bremsstrahlung (free-free) emission dominated by the power-law component of the host galaxy. The absolute oxygen abundance (O/H) of the H {\sc ii} regions that are closest to the seven SNe exploded in $1992-2010$ ranges from $\sim0.53$ to $\sim0.70$ solar \citep{and11}. The absolute oxygen abundance (O/H) of the interstellar medium of NGC\,3690 is $\sim0.66$ solar \citep{hec15}. 
% log(O/H) + 12 = 8.51 to 8.63

In other well studied X-ray bright starburst galaxies, super-solar Ne/O and Mg/O have been reported. For NGC\,253, the Mg/O and Ne/O abundance ratios are $\sim2-3$ times solar for the central and surrounding regions \citep{bau07}. For M82, while the stellar abundances are almost solar, the hot atmosphere Mg/O and Ne/O abundance ratios are $\sim2-5$ times solar \citep{ori04}. After taking into account the charge-exchange effect \citep{zha14}, these two abundance ratios are still about twice solar with Ne/O$=2.2\pm0.3$ and Mg/O$=2.1\pm0.3$, respectively. 

% word count: 313 

\section{Metal enrichment via supernovae}
To interpret the observed abundance ratios, we build a simplified chemical enrichment model based on SNcc and SNIa nucleosynthesis yields. SNIa is included here to take into account the potential contribution of a ``prompt" population of SNIa \citep{maoz14} that explode within a few hundreds of million years after the peak star-formation activity. The theoretical abundance ratios are calculated via 
\begin{equation}
    \frac{z_i}{z_j} = \frac{\bar{y}_i^{\rm SNcc} + r^{\rm SNIa}~y_i^{\rm SNIa}}{\bar{y}_j^{\rm SNcc} + r^{\rm SNIa}~y_{\rm j}^{\rm SNIa}} \frac{A_j~n_j}{A_i~n_i}~,
\end{equation}
where $z_{i,j}$ are the abundance of the $i$th or $j$th element, $y_i^{\rm SNcc}$ the IMF-weighted SNcc yields, $r^{\rm SNIa}$ the number ratio of SNIa with respect to SNcc, $y_i^{\rm SNIa}$ the SNIa yields, $A_{i,j}$ the atomic weight of the $i$th or $j$th element and $n_{i,j}$ are the elemental abundance by number in the (proto-)solar abundance table. 

We do not have accurate abundances of Fe-peak abundances (e.g., Cr, Mn, and Ni) other than Fe. Therefore, we cannot break degeneracy of dozens of SNIa yields available in the literature. For simplicity, we adopt the widely used W7 yields of \citet{iwa99} when investigating the contribution from a ``prompt" population of SNIa.  

The IMF weighted SNcc yields\footnote{Only single-star SN nucleosynthesis yields are considered here.} are calculated via: 
\begin{equation}
\label{eq:piwy}
    \bar{y}_i^{\rm SNcc} = \frac{\int_{m_{\rm lo}}^{m_{\rm up}} \phi(m)~y_i(m)~dm}{\int_{m_{\rm lo}}^{m_{\rm up}} \phi(m)~dm}~,
\end{equation}
where $\phi(m)\propto m^{\Gamma}$ is the IMF, $m_{\rm lo}$ and $m_{\rm up}$ are the lower and upper mass limits of the progenitors, which depends on its initial metallicity. The Salpeter IMF \citep[][$\Gamma=-2.35$]{sal55} is adopted here as the default. SNcc yields are sourced from \citet{nom13} and \citet{suk16}. \citet[][N13]{nom13} provide SNcc yields of massive stars with different initial metallicities ($Z=$0, 0.001, 0.004, 0.008, 0.02, and 0.05). The mass range of the progenitor are $11-140~M_{\odot}$ for $Z=0$ and $13-40~M_{\odot}$ for $Z>0$. Taking advantage of a one dimensional neutrino transport model for the explosion, \citet[][S16]{suk16} provide SNcc yields of massive ($12.25-120~M_{\odot}$) stars with solar metallicity ($Z=0.02$). Both the W18 and N20 models of \citet{suk16} are considered here. The former was calibrated so that a \citet{nom88} progenitor explodes like SN\,1987A, while the latter was calibrated so that a \citet{utr15} progenitor explodes like SN\,1987A. 

\begin{figure*}[!t]
\centering
\includegraphics[width=\hsize, trim={0.5cm 0.5cm 1.cm 0.cm}, clip]{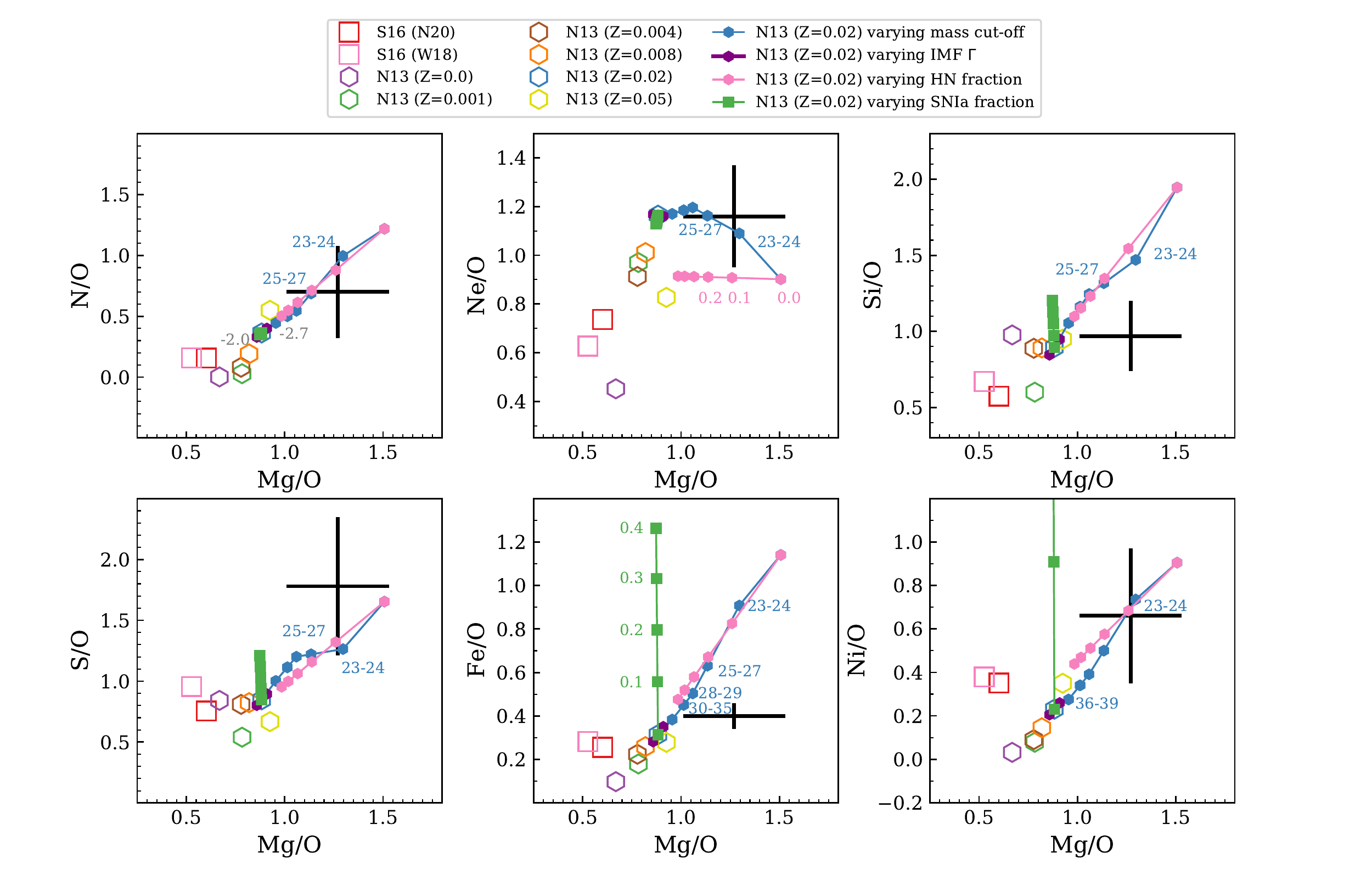}
\caption{Elemental abundance patterns of the hot atmosphere of Arp\,299. The observed abundance ratios (with $1\sigma$ uncertainties) are shown as the black crosses. Chemical enrichment models with core-collapse supernovae (SNcc) yields from \citet[][S16]{suk16} and \citet[][N13]{nom13} are shown as squares and hexagons, respectively. For the N13 models, we further consider different IMF power-law indices $-2.0$ (left) and $-2.7$ (right); different cut-off masses for the SNcc progenitors with the upper mass limit $M_{\rm up}$: $36-39~M_{\odot}$ (leftmost), $30-35~M_{\odot}$, $28-29~M_{\odot}$, $25-27~M_{\odot}$, $23-24~M_{\odot}$, and $20-22~M_{\odot}$ (rightmost); different hypernova fractions from 0.5 (leftmost) to 0.0 (rightmost) for massive stars with $M_{\star}\sim20-40~M_{\odot}$ (and SNcc for stars with $M_{\star}\sim13-20~M_{\odot}$), and different SNIa fractions from 0.0 (bottom) to 0.4 (top).}
\label{fig:cf_arp_obs_all}
\end{figure*}

As shown in Figure~\ref{fig:cf_arp_obs_all}, the calculated IMF weighted abundance ratios are not within the $\gtrsim1\sigma$ uncertainties of the observed ones. The closest match is the one with N13 yields for massive ($13-40~M_{\odot}$) stars with solar metallicity ($Z=0.02$). In the following, we tried to vary several underlying parameters based on the N13 $Z=0.02$ model. 

First, we tried to vary the IMF power-law index. For Arp\,299, the relatively high frequency of ``stripped-envelope" SNe (type Ib and type IIb) with respect to normal type II SNe suggests that the IMF of the host galaxy is biased toward the production of high-mass stars \citep{and11}\footnote{Note that ``stripped-envelope" SNe can also be produced by binary stars with a less massive progenitor \citep{sma09}.}. In Figure~\ref{fig:cf_arp_obs_all}, the purple filled circles illustrate the effect of varying IMF power-law index. A top-heavy IMF with $\Gamma=-2.0$ can decrease the abundance ratios (X/O) by $\lesssim0.06$, while a bottom-heavy IMF with $\Gamma=-2.7$ can increase the abundance ratios (X/O) by $\lesssim0.06$. Both cases do not improve the mismatch. 

Varying the upper mass limit of the massive progenitors can bring the calculated abundance ratios closer to the observed ones. The blue curve and filled circles in Figure~\ref{fig:cf_arp_obs_all} illustrate the effect of varying $m_{\rm up}$ in Eq.~\ref{eq:piwy}. The observed abundance ratios N/O, Ne/O, Mg/O, S/O, and Ni/O can be better explained if massive progenitors up to $\sim23-27~M_{\odot}$ explode as SNcc and enrich the hot atmosphere with metals. The Si/O--Mg/O and Fe/O--Mg/O patterns favor $m_{\rm up}\gtrsim30~M_{\odot}$ with the observed Mg/O ratio larger than the theoretical ones. If the theoretical SNcc nucleosynthesis yields of Mg/O yields are underestimated \citep{gri21}, say by 20\%, a more consistent match might be found for all six abundance patterns. In this case, the upper limit of the SNcc progenitors can be larger than $\sim23-27~M_{\odot}$. 

We further consider a more complicated scenario where massive progenitors with $M_{\star}\sim13-20~M_{\odot}$ explode as SNcc while a fraction of massive progenitors with $M_{\star}\sim20-40~M_{\odot}$ explode as HN\footnote{\citet{nom13} HN yields are available for $M_{\star}=$20, 25, 30, and 40~$M_{\odot}$ with solar metallicity.}. The pink curve and filled circles in Figure~\ref{fig:cf_arp_obs_all} illustrate this mixing effect. The abundance pattern of Ne/O--Mg/O can break the degeneracy between the HN fraction and $m_{\rm up}$ cut-off. Taking into account that SNcc are the dominant metal factory of O, Ne, and Mg (i.e., contributions from AGBs and SNIa are negligible), the Ne/O--Mg/O pattern is the best one to explore the difference of theoretical chemical enrichment models of SNcc. 

We also consider an alternative complicated scenario where massive progenitors with a mixture of SNcc (from massive stars with $M_{\star}\sim13-40~M_{\odot}$) and SNIa. Here we consider the number of SNIa to that of SNcc ($r_{\rm SNIa}$ in Eq.~\ref{eq:piwy}) ranges from 0.0 to 0.4. The green curve and filled squares in Figure~\ref{fig:cf_arp_obs_all} illustrate this mixing effect. The abundance patterns of N/O--Mg/O and Ne/O--Mg/O are barely affected by the SNIa fraction. As shown in \citet{mer16,sim19,mao19}, even for rather old systems like elliptical galaxies, groups and clusters of galaxies, where a significant fraction of SNIa has contributed to the chemical enrichment process, SNcc is still the dominant metal factory of O, Ne, and Mg. For heavier elements including Si, the increasing SNIa fraction will increase the abundance ratios to O. 

In short, we find a hint of deviation between the observed abundance patterns and theoretical ones from a simple chemical enrichment model. The deviation can be explained if massive stars with $M_{\star}\gtrsim23-27~M_{\odot}$ do not explode as SNcc. This is in accordance with the missing massive SNcc progenitors problem, where SNcc observed so far are not produced by massive stars with $M_{\star}\gtrsim18~M_{\odot}$ \citep{sma15}. It is also possible that theoretical SNcc nucleosynthesis yields of Mg/O yields are underestimated \citep{gri21}. Statistical uncertainties of the current data are still large, which can be significantly improved with the next generation of X-ray spectrometer like XRISM/Resolve \citep{xri20} and Athena/X-IFU \citep{bar18}. Last but not least, we caution that the observed abundance ratios might not be directly compared to our simple chemical enrichment models. The metals produced by SNcc are distributed not only in the X-ray emitting hot atmosphere but also in the cooler phase of the interstellar medium (gas and dust). Inflows and outflows can further complicate the distribution and dilution of the metals. 

% word count: 1077 

%\section{Software and third party data repository citations} \label{sec:cite}

%In the bibliography the format for data or code follows this format: \\
%\noindent author year, title, version, publisher, prefix:identifier\\

%\acknowledgments

\begin{acknowledgments}
This work is based on observations obtained with XMM-Newton, an ESA science mission with instruments and contributions directly funded by ESA Member States and NASA. We thank the referee for the careful reading of the manuscript and useful suggestions to improve the quality of this work. JM acknowledges useful discussions with Jiangtao Li and H. Yamaguchi. PZ acknowledges the support from NSFC grant 11590781. AS is supported by the Women In Science Excel (WISE) programme of the Netherlands Organisation for Scientific Research (NWO), and acknowledges the World Premier Research Center Initiative (WPI) and the Kavli IPMU for the continued hospitality. YS acknowledges support from Chandra Grants AR8-19020A and GO1-22126X. SRON is supported financially by NWO, the Netherlands Organization for Scientific Research.  
\end{acknowledgments}


\vspace{5mm}
\facilities{XMM, HST, Spitzer}


\software{SPEX v3.06.00 \citep{kaa20}}
% astropy \citep{ast13}



\appendix

\section{Spectral models}
\label{sct:spec_mo}
A total of eight spectral fittings were performed in the present work. The first three sets (S, T, M0 in Table~\ref{tbl:fit_pars}) in Table~\ref{tbl:fit_pars} adopt different differential emission measure distributions for the hot atmosphere: single-temperature (Model S), two-temperature (Model T), and a multi-temperature (Model M1) model with a Gaussian (log-normal) differential emission measure (DEM) distribution \citep[GDEM,][]{dpl06}, 
\begin{equation}
\label{eq:gdem}
    Y(x) = \frac{Y_0}{\sigma \sqrt{2\pi}} \exp\left(-\frac{(x-x_0)^2}{2\sigma^2}\right)~,
\end{equation}
where $Y$ and $Y_0$ are the emission measures, $x=\log_{10} T$, $x_0=\log_{10}(T_0)$, and $T_0$ is peak temperature of the DEM distribution in units of keV. The single-temperature (i.e., isothermal) model is nested in Eq.~\ref{eq:gdem} with $\sigma=0$. Both the multi-temperature and two-temperature DEM distributions yield significantly better statistics. %Figure~\ref{fig:plot_mdl_01to20} compares Model S versus M1. In general, the power-law component and Fe K lines dominates the hard X-ray band, while the thermal component with prominent metal lines dominate the soft X-ray continuum. 

% \begin{figure*}
% \centering
% \includegraphics[width=\hsize, trim={0.5cm 0.cm 0.5cm .cm}, clip]{plot_mdl_01to20.pdf}
% \caption{Best-fit model components of Arp\,299. The power-law and Fe K lines are shown in purple. For the thermal component, a single-temperature (1T) model is shown in green, while a multi-temperature (GDEM) model is shown in red. Vertical dashed lines in blue mark the key diagnostic emission lines of various elements.}
% \label{fig:plot_mdl_01to20}
% \end{figure*}

Adding another single-temperature cool component to Model M1 can further improve $C$-statistics by $\sim-16$ at the cost of 2 less degrees of freedom (Model M2). The main improvement comes from O {\sc vii}. Setting $\sigma(\log T)$ of this cool component free does not improve the fit at all with $\sigma(\log T)$ remains at zero. With still some residuals for the O {\sc vii} lines, we set the temperature ratio $T_b/T_e$ for the cool component (Model M3). For collisionally ionized equilibrium plasmas, the ionization balance temperature $T_b$ equals the electron temperature $T_e$. A non-unity value can mimic non-equilibrium plasmas. For Model M3, we obtain $T_b/T_e\sim1.6$, where the forbidden to resonant line ratio of He-like O {\sc vii} triplet was boosted to better match the observed data. But $C$-statistics is merely improved by $\sim-0.8$ with one less degree of freedom. This indicates a minor contribution from non-equilibrium plasmas. The charge exchange process can also increase the the forbidden to resonant line ratio of He-like O {\sc vii} triplet, as in M82 \citep{zha14}. Considering the availability of H {\sc i} and molecular gas in Arp\,299 \citep[e.g.,][]{hec99,sli12}, it is possible that charge exchange between the hot and cold gas play a role here. Unfortunately, the quality of the current data set is not sufficient to constrain the free parameters of the possible charge exchange process, such as the interacting atomic or molecular species, collision velocities, and different capture mechanisms to the $l$-subshells. In addition, based on Model M2, we decouple the turbulence velocity ($v_{\rm RMS}$) of the two thermal components in Model M4. This yields the best $C$-statics among all eight models.

For all previous models (S to M4), the photon index of the power-law component was fixed to 1.67 as measured by \citet{ana16} with joint \textit{Chandra} and \textit{NuSTAR} observations. In Model M5, we set the photon index free, which yields $\Gamma\sim1.59$ but $C$-statistics is merely improved by $\sim-0.8$ with one less degree of freedom (compared to Model M2). 

Last but not least, based on Model M2, we couple the Ni and Fe abundances in Model M6. This is driven by the imperfect Ni-L complex atomic data in SPEX v3.06.00. The best-fit results of Models M2 to M6 also provide an estimate of the systematic uncertainties, which are smaller than the statistical uncertainties of the current data. That is to say, the key abundance ratios (Ne/O, Mg/O, and Fe/O) do not vary significantly in these models. 

\begin{deluxetable}{c|cccccccccccccccc}
\tablecaption{Best-fit parameters and statistics. The hot atmosphere is modeled with either a single-temperature (Model S), two-temperature (Model T), or multi-temperature (Models M1$-$M6) differential emission measure plasma model. The power-law (PL) component and Fe K lines are required for the point-sources. The expected $C$-statistics \citep{kaa17} are $1135\pm49$ (total), 1010 (RGS), and 125 (pn). Statistical uncertainties at the 68\% confidence level are provided for the best-fit (Model M4). Frozen and coupled parameters are shown as (f) and (c), respectively.
\label{tbl:fit_pars}}
\tablehead{
\colhead{Model} & \colhead{S} & \colhead{T} & \colhead{M1}  & \colhead{M2} & \colhead{M3} & \colhead{M4}  & \colhead{M5}  & \colhead{M6}   
}
%\colnumbers
\startdata
\noalign{\smallskip} 
\multicolumn{9}{c}{Hot atmosphere} \\
\noalign{\smallskip}
\hline
\noalign{\smallskip} 
$N_{\rm H}^{\rm HA}~(10^{21}~{\rm cm^{-2}})$ & 0.9 & 1.2 & 1.3 & 1.5 & 1.7 & $1.6_{-0.3}^{+0.5}$ & 1.6 & 1.5 \\
\noalign{\smallskip} 
$N_{\rm hot}~(10^{64}~{\rm cm^{-3}})$ & 1.7 & 1.5 & 2.1 & 2.2 & 2.4  & $2.4_{-0.5}^{+0.8}$ & 2.4 & 2.3 \\
\noalign{\smallskip} 
$T_{\rm hot}$ & 0.59 & 0.62 & 0.61 & 0.67 & 0.66 & $0.67\pm0.04$ & 0.68 & 0.68 \\
\noalign{\smallskip} 
$\sigma_{\rm hot} (\log T)$ & \nodata & \nodata & 0.50 & 0.33 & 0.35 & $0.31\pm0.05$ & 0.36 & 0.32 \\
\noalign{\smallskip} 
$v_{\rm RMS,~hot}$ & 830 & 930 & 810 & 840 & 840 & $840\pm60$ & 830 & 840 \\
\noalign{\smallskip} 
$N_{\rm cool}~(10^{64}~{\rm cm^{-3}})$ & \nodata & 0.3 & \nodata & 0.8 & 3.3 & $0.6_{-0.3}^{+2.1}$ & 0.9 & 0.9 \\
\noalign{\smallskip} 
$T_{\rm cool}$ & \nodata & 0.18 & \nodata & 0.11 & 0.086 & $0.12\pm0.04$ & 0.11 & 0.11 \\
\noalign{\smallskip} 
$T_b/T_e (\rm cool)$ & \nodata & 1.0 (f) & \nodata & 1.0 (f) & 1.6 & 1.0 (f) & 1.0 (f) & 1.0 (f) \\
\noalign{\smallskip} 
$v_{\rm RMS,~cool}$ & \nodata & 930 (c) & \nodata & 840 (c) & 840 (c) & $380_{-190}^{+250}$ & 830 (c) & 840 (c) \\
\noalign{\smallskip} 
N/O & 0.85 & 0.58 & 0.87 & 0.69 & 0.53 & $0.70_{-0.29}^{+0.38}$ & 0.70 & 0.68 \\
\noalign{\smallskip} 
Ne/O & 1.21 & 1.47 & 1.44 & 1.22 & 1.20 & $1.16\pm0.21$ & 1.20 & 1.21 \\
\noalign{\smallskip} 
Mg/O & 1.00 & 1.21 & 1.81 & 1.36 & 1.35 & $1.27\pm0.26$ & 1.34 & 1.29 \\
\noalign{\smallskip} 
Si/O & 1.07 & 1.17 & 1.42 & 1.04 & 1.02 & $0.97\pm0.23$ & 1.05 & 0.96 \\  
\noalign{\smallskip} 
S/O & 3.8 & 3.9 & 2.2 & 1.89 & 1.77 & $1.78\pm0.57$ & 1.84 & 1.73 \\
\noalign{\smallskip} 
Fe/O & 0.24 & 0.31 & 0.53 & 0.42 & 0.42 & $0.40_{-0.05}^{+0.06}$ & 0.43 & 0.41 \\
\noalign{\smallskip} 
Ni/O & 0.97 & 1.06 & 1.28 & 0.73 & 0.80 & $0.66_{-0.27}^{+0.31}$ & 0.64 & 0.41 (c) \\
\noalign{\smallskip} 
\hline
\noalign{\smallskip} 
\multicolumn{9}{c}{Point sources} \\
\noalign{\smallskip}
\hline
\noalign{\smallskip} 
$N_{\rm H}^{\rm PS}~(10^{21}~{\rm cm^{-2}})$ & 4.0 & 4.0 & 5.1 & 6.3 & 6.8 & $6.7_{-1.5}^{+3.2}$ & 5.7 & 6.6 \\
\noalign{\smallskip} 
$N_{\rm PL}~(10^{49}~{\rm ph~s^{-1}~keV^{-1}})$ & 7.87 & 7.85 & 7.17 & 8.09 & 7.43 & $7.48\pm0.02$ & 6.56 & 7.43 \\
\noalign{\smallskip} 
$\Gamma_{\rm PL}$ & 1.67 (f) & 1.67 (f) & 1.67 (f) & 1.67 (f) & 1.67 (f) & 1.67 (f) & 1.59 & 1.67 (f) \\
\noalign{\smallskip} 
$N_{6.4~\rm keV}~(10^{47}~{\rm ph~s^{-1}})$ & 11.8 & 11.8 & 12.6 & 12.4 & 12.5 & $12.43\pm0.14$ & 12.1 & 12.5 \\
\noalign{\smallskip} 
$N_{6.7~\rm keV}~(10^{47}~{\rm ph~s^{-1}})$ & 7.1 & 7.1 & 7.2 & 7.3 & 7.3 & $7.3\pm1.3$ & 7.1 & 7.4 \\
\noalign{\smallskip} 
$N_{7.0~\rm keV}~(10^{47}~{\rm ph~s^{-1}})$ & 2.1 & 2.2 & 2.9 & 2.6 & 2.7 & $2.6\pm1.1$ & 2.4 & 2.7 \\
\noalign{\smallskip} 
\hline
\noalign{\smallskip} 
\multicolumn{9}{c}{Statistics} \\
\noalign{\smallskip}
\hline
\noalign{\smallskip} 
$C_{\rm stat}$ (total) & 1420.0 & 1353.2 & 1311.5 & 1295.8 & 1294.6 & 1292.8 & 1294.6 & 1297.1 \\
d.o.f. (total) & 1080 & 1078 & 1079 & 1077 & 1076 & 1076 & 1076 & 1078 \\
$C_{\rm stat}$ (RGS) & 1268.4 & 1203.7 & 1171.7 & 1157.7 & 1156.7 & 1155.1 & 1158.0 & 1157.6 \\
$C_{\rm stat}$ (pn) & 151.6 & 149.4 & 139.8 & 138.1 & 137.9 & 137.7 & 136.6 & 139.5 \\
\noalign{\smallskip} 
\enddata
%\tablenotetext{a}{TBA}
%\tablecomments{TBA}
\end{deluxetable}

\end{document}